\documentclass[secnumarabic,nobibnotes,aps,prl,letterpaper,preprintnumbers,superscriptaddress,floatfix,twocolumn]{revtex4-1}
\usepackage{amssymb}
\usepackage{amsmath}
\usepackage{graphicx}
\usepackage{dcolumn}
\usepackage{bm}
\usepackage{multirow}
\usepackage{lineno}
\usepackage[letterpaper,margin=2cm]{geometry}
\begin{document}

\title{Spin-lattice and electron-phonon coupling in 3$d$/5$d$ hybrid Sr$_3$NiIrO$_6$}

\author{Kenneth R. O'Neal}
\affiliation{Department of Chemistry, University of Tennessee, Knoxville, Tennessee 37996, USA}
\author{Arpita Paul}
\affiliation{Department of Chemical Engineering and Materials Science, University of Minnesota, Minneapolis, MN 55455, USA}
\author{Amal al-Wahish}
\affiliation{Department of Chemistry, University of Tennessee, Knoxville, Tennessee 37996, USA}
\author{Kendall D. Hughey}
\affiliation{Department of Chemistry, University of Tennessee, Knoxville, Tennessee 37996, USA}
\author{Avery L. Blockmon}
\affiliation{Department of Chemistry, University of Tennessee, Knoxville, Tennessee 37996, USA}
\author{Xuan Luo}
\affiliation{Max Planck POSTECH/Korea Research Initiative, Pohang University of Science and Technology, Pohang 37673, Korea}
\affiliation{Key Laboratory of Pohang Emergent Materials, Pohang Accelerator Laboratory, Pohang 37673, Korea}
\affiliation{Key Laboratory of Materials Physics, Institute of Solid State Physics, Chinese Academy of Sciences, Hefei, 230031, China}
\author{Sang-Wook Cheong}
\affiliation{Max Planck POSTECH/Korea Research Initiative, Pohang University of Science and Technology, Pohang 37673, Korea}
\affiliation{Key Laboratory of Pohang Emergent Materials, Pohang Accelerator Laboratory, Pohang 37673, Korea}
\affiliation{Rutgers Center for Emergent Materials and Department of Physics and Astronomy, Rutgers University, Piscataway, New Jersey 08854, USA}
\author{Vivien S. Zapf}
\affiliation{National High Magnetic Field Laboratory, MS E536, Los Alamos National Laboratory, New Mexico, 87545, USA}
\author{Craig V. Topping}
\affiliation{School of Physics \& Astronomy, University of St Andrews, North Haugh, St Andrews, KY16 9SS, United Kingdom}
\author{John Singleton}
\affiliation{National High Magnetic Field Laboratory, MS E536, Los Alamos National Laboratory, New Mexico, 87545, USA}
\author{Mykhalo Ozerov}
\affiliation{National High Magnetic Field Laboratory, Tallahassee, Florida 32310, USA}
\author{Turan Birol}
\affiliation{Chemical Engineering and Materials Science, University of Minnesota, Minneapolis, MN 55455, USA}
\author{Janice L. Musfeldt}
\affiliation{Department of Chemistry, University of Tennessee, Knoxville, Tennessee 37996, USA}
\affiliation{Department of Physics and Astronomy, University of Tennessee, Knoxville, Tennessee 37996, USA}
\email{musfeldt@utk.edu}

\date{\today}

%\begin{document}

%\linenumbers

%%%%%%%%%%%%%%%%%%%%%%%%%%%%%%%%
\begin{abstract}
%%%%%%%%%%%%%%%%%%%%%%%%%%%%%%%%

{\bf While 3$d$-containing materials display strong electron correlations, narrow band widths, and robust magnetism, 5$d$ systems are recognized for strong spin-orbit coupling, increased hybridization, and more diffuse orbitals. Combining these properties leads to novel behavior.  Sr$_3$NiIrO$_6$, for example, displays complex magnetism and ultra-high coercive fields - up to an incredible 55~T. Here, we combine infrared and optical spectroscopies with high-field magnetization and first principles calculations to explore the fundamental excitations of the lattice and related coupling processes including spin-lattice and electron-phonon mechanisms. Magneto-infrared spectroscopy reveals spin-lattice coupling of three phonons that modulate the Ir environment to reduce the energy required to modify the spin arrangement. While these modes primarily affect exchange within the chains, analysis also uncovers important inter-chain motion. This provides a mechanism by which inter-chain interactions can occur in the developing model for ultra-high coercivity. At the same time, analysis of the on-site Ir$^{4+}$ excitations reveals vibronic coupling and extremely large crystal field parameters that lead to a t$_{2g}$-derived low-spin state for Ir. These findings highlight the spin-charge-lattice entanglement in Sr$_3$NiIrO$_6$ and suggest that similar interactions may take place in other 3$d$/5$d$ hybrids.}

\end{abstract}

\maketitle

%%%%%%%%%%%%%%%%%%%%%%%%%%%%%%%%
\section*{Introduction}
%%%%%%%%%%%%%%%%%%%%%%%%%%%%%%%%

Interest in 5$d$ materials and 3$d$/5$d$ hybrids has blossomed in response to recent scientific advances and applications in hard magnets, topological insulators, multiferroics, superconductors, and thermoelectrics~\cite{Rau2016,Witczak-Krempa2014,Feng2017,Feng2018,Liu2019,Isobe2019}. 5$d$ materials are unique for several reasons. First, strong spin-orbit coupling competes with magnetic, crystal-field, many-body Coulomb, and other interactions to drive new physical behaviors  \cite{Cao2018}, such as the $J_{\rm eff}=1/2$ state in certain iridates~\cite{Kim2009,Clancy2018}.
%First, the SOC tends to be much stronger, leading to the possibility of band inversion, strong Rashba and spin texture effects. However, crystal field effects also tend to be stronger, bandwidths are larger, and many-body Coulomb interactions are effectively weaker compared to materials with more localized 3d and 4d electrons. The competition between these interactions therefore plays out different ways and drives new physical behaviors, such as the Jeff = 1/2 state that emerges in certain iridates.
Second, the bonding interactions associated with the larger 5$d$ orbitals promote inter-cation dimerization in pairwise, chain-like, and other complex orderings~\cite{Yang2012a,Hsu2013}. Third, the relativistic shifts in orbital energies, combined with spin-orbit and bandwidth effects, can drive band inversions leading to topological phases and enhanced Rashba splittings~\cite{Zheng2015,Goldman2011,Bihlmayer2007,Sie2019a}. In contrast, 3$d$ transition metal compounds typically display much narrower bandwidths, more robust magnetism, and stronger electron-electron interactions, and correlations~\cite{Charnukha2017,Liu2018}. When these two sets of properties are brought together, as in Sr$_3$NiIrO$_6$, new and potentially useful behaviors can emerge.

What makes Sr$_3$NiIrO$_6$ so remarkable %compared with other hard magnets 
is the extraordinary coercivity - up to 55~T depending on sample details \cite{Singleton2016}. By contrast, traditional hard magnets like Fe/Pt, Nd$_{31-x}$Dy$_x$Fe$_{\rm bal}$Co$_2$B$_1$ ($x$ = 7 wt \%), and LuFe$_2$O$_4$ have coercivities on the order of 1, 3, and 9~T, respectively~\cite{Ristau1999,Bai2007,Wu2008}. The extraordinarily high coercive field is not due to ferromagnetic domains since the material is antiferromagnetic, though the exact mechanism remains to be understood~\cite{Singleton2016,Ou2014,Zhang2010,Birol2018}.
%While single ion anisotropy is the most common origin of a large coercive field,  recent calculations support the role of anisotropic exchange interactions in Sr$_3$NiIrO$_6$~\cite{Birol2018}. 

%%% Vivien's suggestion for introducing the magnetic order of SNIO and the puzzle of the high coercive field:
The crystal structure consists of chains of alternating face-sharing NiO$_6$ trigonal prisms and IrO$_6$ octahedra stacked along the $c$ axis with Sr atoms separating the chains. These chains are arranged in a triangular configuration in the $ab$ plane~\cite{Mikhailova2012}. Because Ir has a 4+ charge, it was originally thought to have a $J_{\rm eff} = 1/2$ state~\cite{Lefrancois2014,Toth2016,Zhang2010}. A trigonal distortion, however, breaks the symmetry and creates a slight deviation from the pure $J_{\rm  eff} = 1/2$~\cite{Lefrancois2016}. 
Sr$_3$NiIrO$_6$ is part of a chemical family of quasi-one-dimensional 3$d$/5$d$ hybrids with formula $A_3BB'$O$_6$. Notably, these materials display increasing coercive magnetic fields as the $B'$ site ion evolves from a 3$d$ to a 4$d$ to a 5$d$ magnetic ion~\cite{Mikhailova2012,Stitzer2002,Singleton2016}. This suggests that the unusual properties of Ir and similar 5$d$ ions must play an important role in creating the high coercivity. Recent calculations indeed suggest that it is the strongly anisotropic exchange interaction characteristic of the Ir$^{4+}$ ion that plays a key role the high coercive magnetic field~\cite{Birol2018}. %Elastic neutron diffraction measurements at zero magnetic field identify two possible magnetic ground states below 75~K: a partially disordered antiferromagnet (PDA) or a more complex spatially modulated state.~\cite{Lefrancois2014}. Both are predominately antiferromagnetic and must evolve in magnetic fields to create the magnetic hysteresis with high coercive field. 
The strongest magnetic exchange is antiferomagnetic between Ni and Ir ions within $c$-axis chains, which creates one-dimensional order below $\approx$200 K~\cite{Toth2016}. %Thus a likely scenario is that chains remain intact up to 65 T, while frustrated inter-chain ordering drives the magnetic behavior between 0 and 65 T including the high coercivity~\cite{Lefrancois2014,Singleton2016}. 
Elastic neutron diffraction measurements identified two potential magnetic ground states at zero field: a partially disordered antiferromagnet or a more complex, spatially modulated state~\cite{Lefrancois2014}. In light of the large Ni$\cdots$Ir intrachain exchange interactions~\cite{Toth2016}, the partially disordered antiferromagnet model is most likely %since it preserves the intrachain magnetic structure. 
and has been suggested for several $A_3BB'$O$_6$ materials with high coercive field \cite{Niitaka2001,Mohapatra2007,Murthy2019}. In this model, $c$-axis chains act like giant spin units due to the difference between the Ni and Ir moments, with frustrated antiferromagnetic interactions between chains in the $ab$-plane. Consequently, the ground state consists of two oppositely aligned chains and one with random orientation. 
%For Sr$_3$NiIrO$_6$, the chains have a net magnetic moment due to the difference between the moment of the antiferromagnetically aligned Ir and Ni spins. 
In this model, gradual flipping of the randomly oriented third chains accounts for the spin glass-like dynamics with magnetic hysteresis and coercivity that is seen below 15~K in Sr$_3$NiIrO$_6$, and for the initial evolution of the magnetization up to the coercive field. At the coercive field, the second chain flips, leading to a sudden jump and the high coercive field. The extraodrinary coercivity must result from the barrier to flipping of the second chain due to its many-body nature and interactions with the lattice including the anisotropic exchange interaction of Ir. Thus if we are to achieve a complete understanding of the extraordinary coercive magnetic field in Sr$_3$NiIrO$_6$ we need to quantify how the magnetic order and Ir electronic state interact with the lattice.

%At the same time, spin-orbit coupling strengthens interactions between the charge, spin, and lattice. Magnetoelectric coupling, for instance, is significant in many 5$d$-containing materials~\cite{Granado1999,Sohn2017,Cao2018,Li2017}. 
%Other interactions in Sr$_3$NiIrO$_6$, such as cross-coupling of fundamental lattice excitations with the magnetic and electronic responses, are highly under-explored and therefore constitute an important gap in the understanding of strong spin-orbit coupled oxides.%~\cite{Birol2018,Calder2015}  

%\textbf{This constitutes an important gap in the knowledge of magnetism of not only Sr$_3$NiIrO$_6$, but in that of strong SOC oxides in generals, since the so-called \textit{spin-phonon coupling} \cite{Birol2018} can be excessively large in 5d compounds \cite{Calder2015}.} 

To this end, we measured the infrared and optical properties of Sr$_3$NiIrO$_6$ and compared the response to high field magnetization, first principles lattice dynamics calculations, and simulations of  spin-phonon coupling. While the phonons are nearly rigid across the magnetic ordering and freezing transitions, magneto-infrared work reveals three modes that track the magnetic response. What distinguishes these modes is how they modulate superexchange pathways around the Ir centers. The O-Ir-O bending mode near 310~cm$^{-1}$ is the dominant feature with an exceptionally large spin-lattice coupling constant - on the order of 10~cm$^{-1}$. That very specific local lattice distortions are involved in the approach to the coercive field provides further confirmation that the high field transition in Sr$_3$NiIrO$_6$  is more than a simple domain reorientation process \cite{Singleton2016}. At the same time, we reached beyond spin-lattice interactions to explore electron-phonon coupling. Analysis of the Ir$^{4+}$ on-site $d$-to-$d$ excitations reveals that they are vibronically activated. An oscillator strength analysis uncovers electron-phonon coupling involving the 177~cm$^{-1}$ Ni in-plane mode which modulates the Ni-O-Ir bond angles (and thus the local Ir$^{4+}$ environments). We discuss these findings in terms of inter- and intra-chain coupling as well as how the diffuse character of the Ir 5$d$ orbitals enlarges the crystal field parameters to give a spin configuration that derives from the $t_{2g}$-derived energy levels.

%%%%%%%%%%%%%%%%%%%%%%%%%%%%%%%%
\section*{Results and discussion}
%%%%%%%%%%%%%%%%%%%%%%%%%%%%%%%%

\subsection*{Strong spin-lattice coupling and intra-chain interactions in Sr$_3$NiIrO$_6$}

\begin{figure*}[tbh]
\centering
\includegraphics[width=6in]{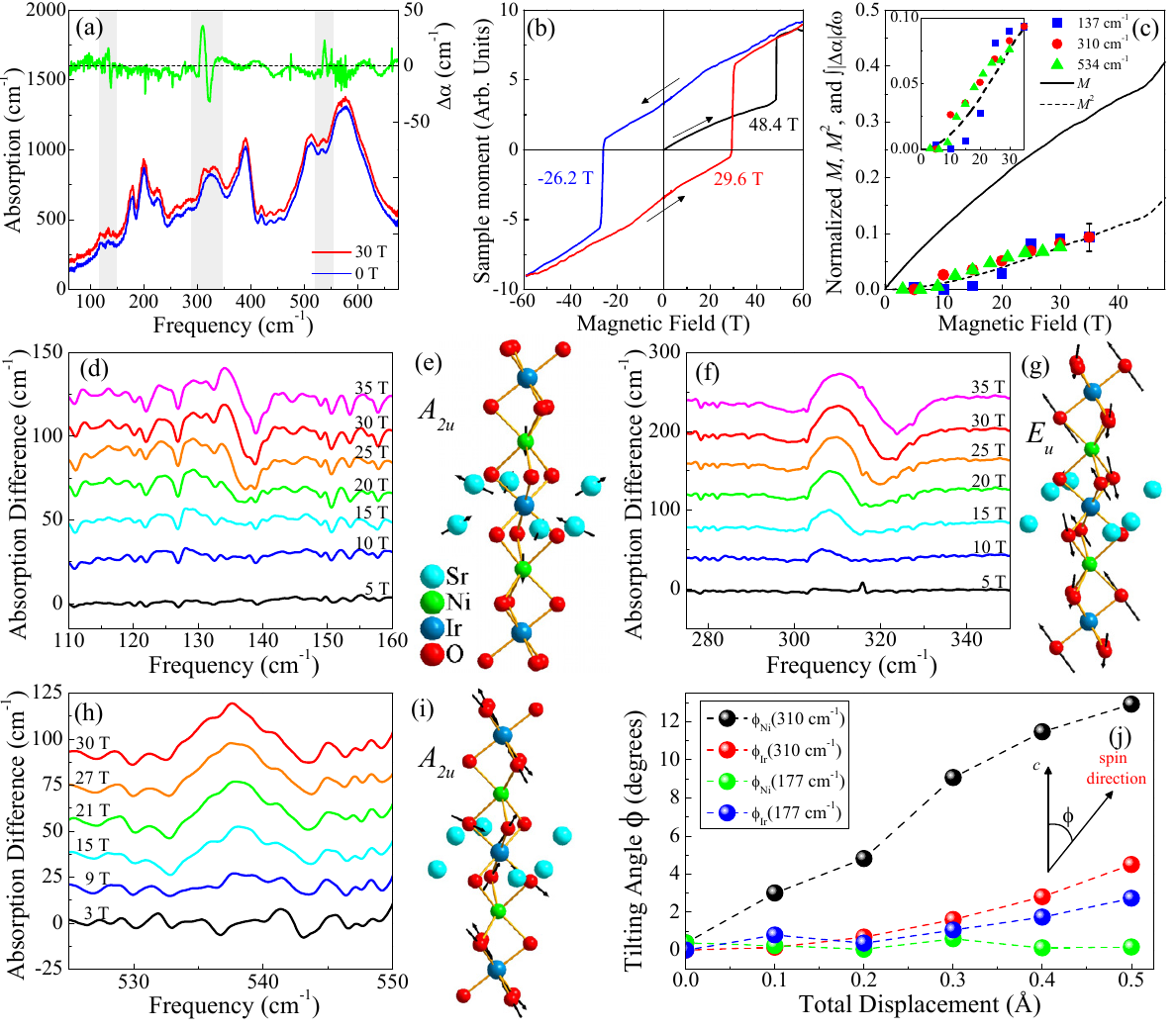}
\caption{ {\bf Magneto-elastic coupling in Sr$_3$NiIrO$_6$.} (a) Absorption of polycrystalline Sr$_3$NiIrO$_6$ at 0 and 30~T at 4.2~K. The high field spectrum is offset for clarity. Top portion displays the 30~T field-induced absorption difference. Vertical gray bands highlight changes. (b) Hysteresis loop of polycrystalline Sr$_3$NiIrO$_6$ at 4.0~K. The indicated critical fields for spin rearrangements are not as high as single crystal measurements due to an averaged response of different orientations. (c) Magnetization, square of the magnetization, and integrated absorption differences for the three features of interest versus magnetic field. Values are normalized at full field for comparison, and a representative error bar (standard deviation) is shown. The inset is a close-up view of the same data. (d,f,h) Close-up views of the absorption difference spectra in the regions of interest and their development with magnetic field. Curves are offset for clarity. (e,g,i) Calculated phonon displacement patterns for these field-sensitive modes. (j) First principles simulation of spin-lattice coupling-induced canting of the Ni and Ir spin moments  with respect to the $c$ axis as a function of ionic displacement chosen to mimic the 310 and 177 cm$^{-1}$ mode patterns. The magnitude of the displacement is given for a total of 22 atoms in the  unit cell (which has two formula units).
\label{SNIO_Field}}
\end{figure*}

Figure \ref{SNIO_Field} (a) displays the infrared spectrum of Sr$_3$NiIrO$_6$. A symmetry analysis of the $R\bar{3}c$ space group yields seven $A_{2u}$ and twelve $E_u$ infrared-active phonons - consistent with our spectra. Our mode assignments are based upon lattice dynamics calculations which also provide symmetries and displacement patterns. These assignments are summarized in Table S1 (Supplementary Information) and allow the microscopic interactions indicated by the spectra to be fully understood. Spin-lattice coupling across the 75~K magnetic ordering and 15~K freezing temperatures is extremely weak (Supplementary Information), likely because these energy scales do not correspond to collective transitions~\cite{Mikhailova2012,Lefrancois2014,Singleton2016}. The signature in other thermodynamic probes is not strong either. Magnetoelastic coupling in applied field, however, is robust.

In order to search for elastic contributions to the record high coercivity, we measured the magneto-infrared response of Sr$_3$NiIrO$_6$ [Fig.~\ref{SNIO_Field}]. Here, magnetic field was increased step-wise from 0 to 35~T, and  spectra were acquired at each step to avoid complications from magnetic hysteresis. The measurements therefore follow the line from the origin in Fig.~\ref{SNIO_Field} (b), shown more closely in Fig.~\ref{SNIO_Field} (c). The absorption difference ($\Delta\alpha=\alpha(30$~T$)-\alpha(0$~T)) in the upper portion of Fig.~\ref{SNIO_Field} (a) highlights spectral differences. Three phonons are sensitive to magnetic field. The E$_u$ symmetry mode at 310~cm$^{-1}$ displays the largest magneto-infrared response  [Fig.~\ref{SNIO_Field} (f)]. The A$_{2u}$ symmetry modes at 133 and 534 cm$^{-1}$ are also sensitive to changes in the microscopic spin arrangement [Fig.~\ref{SNIO_Field}~(d,h)]. Interestingly, the two lower frequency phonons harden with applied field, whereas  the higher frequency mode softens. 

We quantify the magnetic field dependence of the 133, 310, and 534~cm$^{-1}$ phonons by integrating the absolute value of the absorption difference over a narrow frequency window ($\int_{\omega_1}^{\omega_2}|\Delta\alpha|d\omega$) at each field. This quantity is proportional to the field-induced frequency shift, although error bars on ${\Delta}{\alpha}$ are much smaller~\cite{Brinzari2012}. Comparison with the bulk magnetization [Fig.~\ref{SNIO_Field}~(c)] reveals that these changes grow as magnetization squared~\cite{Granado1999}. This implies that the field-driven transition is not just a spin reorientation process; it also involves specific local lattice distortions. The magnetic field drives the spin rearrangement, while cooperative lattice distortions reduce the exchange interactions and the energy required to modify the microscopic spin arrangement. Similar mechanisms are active in other materials~\cite{Granado1999,Poirier2007,Poirier2011}. While the 48~T spin-reorientation field is beyond the current reach of resistive magnets, the observed spin-phonon coupling is quite strong. % - even on approach to the spin-reorientation. %the spin-phonon coupling that takes place on approach to this spin-reorientation transition suggests that  the critical field suggests that additional effects may take place at that spin-reorientation and also across the quantum phase transition to the fully polarized state.%Faster time-scale techniques in pulsed fields, such as magnetostriction, should reveal such effects and shed further light on the underlying mechanism of the large coercive fields. %Using experimental frequency shifts at 30 T and the limiting low temperature value of the spin-spin correlation function \cite{Casto2015}, 
We can estimate the spin-phonon coupling constants in Sr$_3$NiIrO$_6$  as $\omega=\omega_0+\lambda<S_i\cdot S_j>$. Here,  $\omega_0$ is the unperturbed (zero field) mode frequency, $\omega$ is the perturbed (high field) mode frequency, 
and  $<S_i\cdot S_j>$ is the spin-spin correlation function~\cite{Casto2015}.  Taking the limiting low temperature value of the spin-spin correlation function as  $<S_i \cdot S_j>=<{1\cdot \frac{1}{2}}> = \frac{1}{2}$, we find $\lambda\approx 2$, 10, and 5~cm$^{-1}$ for the 133, 310, and 534~cm$^{-1}$ modes, respectively.  By comparison, the highest value of $\lambda$ in the 3$d$/4$p$ hybrid Ni$_3$TeO$_6$ at the same applied magnetic field (i.e. across the spin-flop transition) is 4~cm$^{-1}$~\cite{Yokosuk2015}. This suggests that higher fields may drive even stronger spin-lattice interactions, especially across the coercive field~\cite{Kim2015,Jaime2012}.

Examination of the calculated displacement patterns %provides mechanistic insight and 
uncovers important similarities in the spin-lattice coupled modes that provide mechanistic insight into the field-driven process. The E$_u$ symmetry O-Ir-O bending mode at 310~cm$^{-1}$ is the most sensitive to field [Fig.~\ref{SNIO_Field}~(g)]. %Strong O motion dominates the motion [Fig. \ref{SNIO_Field} (g)]. 
This displacement pattern - with its strong O component - unambiguously modulates the superexchange interactions between Ir and its neighbors by changing the Ir environment. This motion also  reduces the symmetry of the crystal and (as discussed below) introduces new terms into the magnetic Hamiltonian. %Our calculations show that when atoms are displaced according to this pattern, magnetic moments of Ni and Ir are no longer collinear and are tilted by significant amounts (Fig.~\ref{SNIO_Field}~(f)).
Local lattice distortions of this type are thus important to the development of various spin rearrangements - such as that across the coercive field. The other two magneto-infrared-active features modulate the exchange interactions around the Ir centers as well - although by less effective routes. For instance, the A$_{2u}$ symmetry mode at 133~cm$^{-1}$ consists of Sr out-of-plane displacement and in-phase Ni displacement along the $c$ axis [Fig.~\ref{SNIO_Field}~(e)] that slightly varies the Ni-O-Ir angle as a second-order effect. The A$_{2u}$ symmetry mode at 534~cm$^{-1}$, on the other hand, consists mainly of O stretching around the Ir center [Fig.~\ref{SNIO_Field}~(i)] which also impacts the superexchange angles. The aforementioned displacement patterns primarily affect exchange within the chains. A view along the chains (see images and animations of the displacement patterns in Supplementary Information) reveals that these modes have inter-chain motion as well. This supports the role of inter-chain interactions in the developing magnetic model~\cite{Lefrancois2014,Singleton2016} and provides a mechanism by which such effects can occur. %Animations of these phonons are available in Supplementary Information. 

The form of the spin-phonon Hamiltonian for Sr$_3$NiIrO$_6$ is very complicated - even when considering only intra-chain interactions - because various phonons break lattice symmetries and allow novel magnetic interaction terms (such as Dzyaloshinskii-Moriya) to emerge. (See the supplementary information for details.) This makes the first principles calculation of every spin-phonon coupling parameter  practically impossible. We therefore  developed a simplified approach to predict which phonon modes have a stronger effect on the magnetic structure (and vice versa). We began by considering the displacement pattern of the 310~cm$^{-1}$ mode and calculated the effect of this type of modulation on the ground state spin arrangement. Similar calculations using the 177~cm$^{-1}$ mode pattern are included for comparison. As a reminder, the E$_u$ symmetry O-Ir-O bending mode at 310 cm$^{-1}$ engages in spin-lattice coupling, whereas (as we shall see below) the E$_u$ symmetry mode at 177 cm$^{-1}$ 
is the phonon that vibronically activates on-site $d$-to-$d$ excitations of Ir$^{4+}$. When the atoms are not displaced, the Ni and Ir moments are predicted to be collinear and parallel to the $c$ axis (within error bars), consistent with the observed magnetism. Displacing atoms according to the pattern of the 310~cm$^{-1}$ mode leads to very significant tilting of the Ni spin moment as well as weaker tilting of the Ir spin  [Fig.~\ref{SNIO_Field}~(j)]. As a result, the system becomes non-collinear. The 177~cm$^{-1}$ mode, on the other hand, leads to a much smaller effect on the magnetic structure, in line with a lack of field-induced changes of this mode. This reveals the complex nature of the spin-lattice coupling and suggests that the induced tilting may lower the magnetic switching barrier. 

%Our calculations show that when atoms are displaced according to this pattern, magnetic moments of Ni and Ir are no longer collinear and are tilted by significant amounts [Fig.~\ref{SNIO_Field}~(j)]. 
%{\bf This reveals that the strong effect of $E_u$ modes on the magnetic interactions  cannot be quantified by a single parameter such as $\lambda \sim {\partial}^2J/{\partial}u^2$ where $J$ is the exchange that is modulated by the lattice distortion of interest ($u$).  
%It is very likely that this particularly large tilting induced by the combination of spin-orbit and spin-phonon couplings lowers the magnetic switching barrier. }

%%%%%%%%%%%%%%%%%%%%%%%%%%%%%%%%%%%%%%%%%%%%%%%%%%%%%%%%%%%%
\subsection*{Electron-phonon coupling and strength of the crystal field interactions in Sr$_3$NiIrO$_6$}
%%%%%%%%%%%%%%%%%%%%%%%%%%%%%%%%%%%%%%%%%%%%%%%%%%%%%%%%%%%%

\begin{figure*}[tbh]
\centering
\includegraphics[width=6.5in]{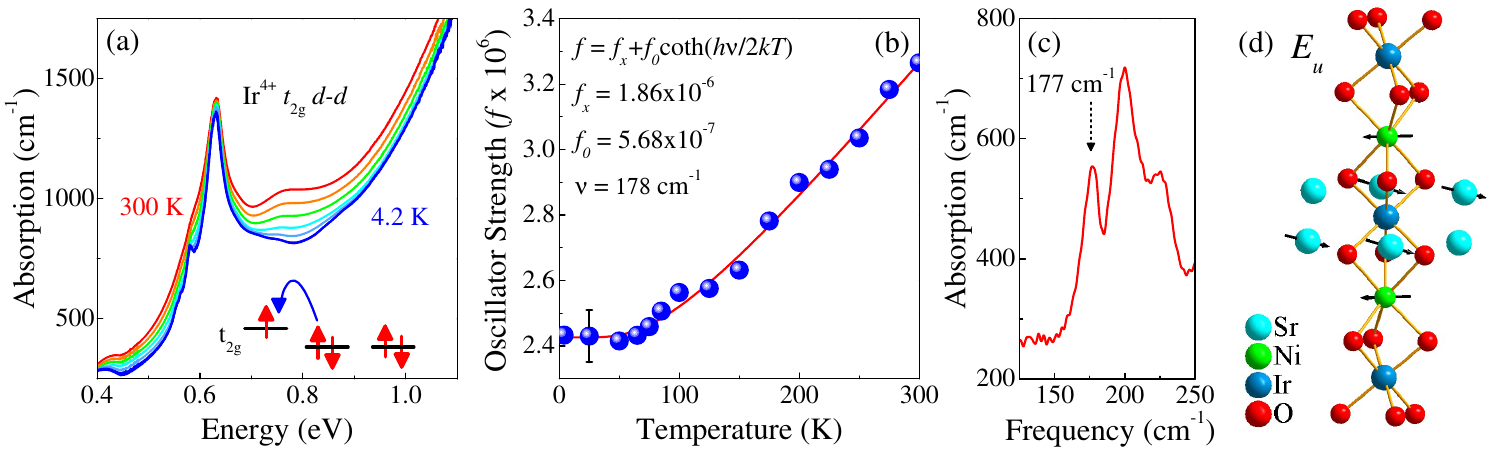}
\caption{ {\bf Vibronic coupling of Ir$^{4+}$ excitations in Sr$_3$NiIrO$_6$.} (a) Optical absorption of Sr$_3$NiIrO$_6$ in the vicinity of the Ir$^{4+}$ on-site excitations at select temperatures. (b) Oscillator strength analysis using the indicated vibronic coupling model, where $\nu$ is the coupled phonon frequency. (c) The infrared spectrum shows phonons near the extracted $\nu$ value. (d) Calculated displacement pattern for the vibronically-coupled phonon.
\label{SNIO_Vib}} \end{figure*}

% Comment form Vivien: Originally this said "Equipped with a microscopic understanding of the elastic contributions to the record-high coercivity" I don't think we have such an understanding. how about we say:
Equipped with a microscopic understanding of the elastic distortions in response to the magnetic field, we sought to determine whether similar phonons contribute to the electronic properties - specifically the vibronically-activated crystal field excitations of Ir$^{4+}$. The Ir$^{4+}$ electronic configuration is key to creating the anisotropic exchange interactions that have been identified as an important contributor to the high coercive field, suggesting that additional insight would prove useful. 
%thus we should track the evolution of this electronic configuration with magnetic field. 
We therefore measured the optical properties of Sr$_3$NiIrO$_6$ as well as the Cu analog, Sr$_3$CuIrO$_6$, to track the Ir$^{4+}$ on-site excitations near 0.7~eV [Fig.~\ref{SNIO_Vib} (a) and Supplementary Information]. While the position and general behavior of these excitations agree with prior resonant inelastic x-ray scattering work~\cite{Lefrancois2016,Liu2012a}, our measurements offer significantly higher spectral resolution. %The $d$-to-$d$ excitations are clustered together in the Ni system, whereas there is some separation in the Cu analog - probably a consequence of the additional chain distortion. In any case, 
The presence of  intra-$t_{2g}$ on-site excitations indicates that the Ir$^{4+}$ site symmetry is slightly distorted from octahedral, creating a distortion away from a pure $J_{\rm eff}=1/2$ state~\cite{Lefrancois2016}. Moreover, analysis reveals that these excitations are vibronically activated by a phonon that is distinct from those that contribute to the coercivity via spin-lattice pathways.
%coupled and that the phonon that activates this process is distinct from those that contribute to the coercivity. %change in magnetic field.

Inter-band excitations, like those in Sr$_3$NiIrO$_6$, are responsible for the colors of transition metal-containing materials and have been extensively studied~\cite{Sell1967,Lohr1972,Ballhausen1962}. 
%These intra-band transitions are formally ``forbidden" but can become ``allowed" through various mixing processes \cite{Sell1967,Lohr1972,Ballhausen1962}. 
Vibronic coupling, in which an odd-parity phonon mixes with a $d$-to-$d$ excitation, is a common activation mechanism~\cite{Sell1967,Lohr1972,Ballhausen1962}. %for activating these excitations 
Here, both spin and parity selection rules are broken due to coupling with a phonon. In this scenario, the temperature dependence of the oscillator strength is modeled as $f=f_x+f_0 \coth (h\nu/2k_{\rm B} T)$, where $\nu$ is the frequency of the activating phonon, $f_0$ is the  oscillator strength at base temperature, $f_x$ represents oscillator strength from other mixing processes, and $h$, $k_{\rm B}$, and $T$ have their usual meanings~\cite{ONeal2017}. This model can be used to determine which phonon activates the electronic transition. While phonon-assisted $d$-to-$d$ excitations have been reported in Sr$_3$Ir$_2$O$_7$~\cite{Park2014}, quantitative vibronic coupling analyses such as we do here are rare in 4- and 5$d$- systems and particularly so in 3$d$/5$d$ hybrids.

Figure~\ref{SNIO_Vib}(b) displays the oscillator strength analysis of the Ir$^{4+}$ on-site excitations in Sr$_3$NiIrO$_6$. There is a small deviation from the overall trend near 70~K that may be due to the spin ordering transition~\cite{Flahaut2003}, but the size of our error bars precludes a detailed analysis. Examination reveals that the Ir-related $d$-to-$d$ excitations are vibronically coupled with the 177~cm$^{-1}$ E$_u$ phonon - which is present in the infrared spectrum [Fig.~\ref{SNIO_Vib}~(c)]. According to our calculations, this mode consists of Ni in-plane motion against Sr counter-motion that indirectly modulates the Ir environment [Fig.~\ref{SNIO_Vib}~(d)]. Thus, the vibronically coupled mode is separate and distinct from those involved in magnetoelastic coupling (133, 310, and 534~cm$^{-1}$). This separation offers the possibility of selective property control. We also carried out the same analysis for the more distorted Cu analog and unveiled coupling to a different displacement in which O motion more directly affects the Ir environment (Supplementary Information). This contrast likely emanates from the dissimilar chain configurations (linear for the Ni system and zigzag for the Cu analog), highlighting the importance of local symmetry in coupling processes.

\begin{table*}
\centering
\caption{Summary of 10$Dq$ and Racah parameters for the 3$d$/5$d$ materials in this study compared to the crystal field parameters of other Ir-containing compounds along with information about the vibronically coupled phonons. No vibronic coupling analysis was found for Sr$_2$IrO$_4$ or Li$_2$IrO$_3$, and the Ni$^{2+}$ excitations in Ni$_3$TeO$_6$ are not vibronically activated, so no phonons are implicated.}
\footnotesize{
\begin{tabular}{|c|c|c|c|c|c|c|c|} \hline
\multirow{2}{*}{Material} & \multirow{2}{*}{Element} & Electronic & \multirow{2}{*}{10$Dq$ (eV)} & \multirow{2}{*}{$B$ (eV)} & Coupled phonon & \multirow{2}{*}{Displacement} & \multirow{2}{*}{Refs.} \\
         &         & state      &             &          & frequency (cm$^{-1}$) & & \\ \hline
Sr$_3$NiIrO$_6$      & Ir & 5$d^5$ & 3.24 & 1.18 & 177 & Ni in-plane motion & This work \\ \hline
Sr$_3$CuIrO$_6$      & Ir & 5$d^5$ & 2.33 & 0.86 & 273 & O-Ir-O bend & This work \\ \hline
Sr$_2$IrO$_4$        & Ir & 5$d^5$ & 3.8  & 0.93 & -   & - & \citenum{Sala2014,Ishii2011} \\ \hline
Li$_2$IrO$_3$        & Ir & 5$d^5$ & 2.7  & 0.95 & -   & - & \citenum{Gretarsson2013} \\ \hline
Ni$_3$TeO$_6$        & Ni & 3$d^8$ & 1.10 & 0.11 & -   & - & \citenum{Yokosuk2016} \\ \hline
$\alpha$-Fe$_2$O$_3$ & Fe & 3$d^5$ & 1.59 & 0.09 & 525 & In-plane, in-phase Fe-O stretch & \citenum{Marusak1980,ONeal2014a} \\ \hline
\end{tabular}}
\label{SNIO_Racah}
\end{table*}

The observation of Ir$^{4+}$ excitations offers an opportunity to compare crystal field parameters of a 5$d$ center to the more commonly studied 3$d$-containing oxides. Based on the position and shape of the Ir-related intra-band excitations of Sr$_3$NiIrO$_6$~\cite{Lefrancois2016} and the $d^5$ Tanabe-Sugano diagram~\cite{Tanabe1954,Tanabe1954b,Tanabe1956}, we estimate 10$D$q = 3.24~eV and the Racah parameter $B$ = 1.18~eV. As a reminder, 10$Dq$ and $B$ describe the strength of crystal field interactions, and because 5$d$ orbitals are highly diffuse, the crystal field parameters are large. We find 10$Dq$ = 2.33~eV and $B$ = 0.86~eV for Sr$_3$CuIrO$_6$~\cite{Liu2012a}. As summarized in Table~1, these values are similar to other Ir$^{4+}$-containing materials including Sr$_2$IrO$_4$ and Li$_2$IrO$_3$~\cite{Sala2014,Ishii2011,Gretarsson2013} but much higher than prototypical transition metal oxides like $\alpha$-Fe$_2$O$_3$~\cite{Marusak1980} and even the 3$d$/4$p$ hybrid Ni$_3$TeO$_6$~\cite{Yokosuk2016}. We attribute this difference to the heavy mass of the Ir center, which is predicted to increase spin-orbit coupling and Racah parameters in free ions~\cite{Ma2014}. Importantly, the large 10$Dq$ values separate the $t_{2g}$ and $e_g$ levels such that the $e_g$-derived bands in Sr$_3$NiIrO$_6$ play no role in determining the Ir spin configuration.

In summary, we combined infrared and optical spectroscopy with high field magnetization and first principles calculations to explore coupling processes involving the fundamental excitations of the lattice in Sr$_3$NiIrO$_6$ - a material with significant spin-orbit interactions. These include both spin-lattice  and electron-phonon processes.  
%to probe the quasi-$J_{eff}$=1/2 state of the Ir$^{4+}$ ion. 
Magneto-infrared spectroscopy reveals that three phonons - all of which modulate the magnetic pathways around and the symmetry of the Ir centers - display strong spin-lattice interactions, demonstrating that the approach to the coercive field takes place with very specific local lattice distortions - different from expectations for simple domain reorientation in a ferromagnet. 
%implying that the large coercive field stems from more than simple domain physics and that the spin reorientation takes place with lattice involvement. 
Examination of the  mode displacement patterns also provides a specific mechanism for  inter-chain interactions, a finding that is crucial to the development of the working magnetic model in Sr$_3$NiIrO$_6$ and related materials. At the same time, analysis of the on-site Ir$^{4+}$ excitations unveils vibronic coupling and extremely large crystal field parameters. For instance, 10$Dq$ is a factor of two larger than that in traditional transition metal oxides, and the Racah parameter $B$ is a factor of 10 higher. 
Moreover, the phonon that activates the vibronic coupling has a completely different displacement pattern than those that are sensitive to magnetic field. We therefore find that certain phonons in Sr$_3$NiIrO$_6$ are strongly entangled with the spin and charge channels. In addition, the diffuse character of the Ir 5$d$ orbitals determines the ground state spin structure of Sr$_3$NiIrO$_6$ whereas spin-lattice interactions reduce the energy required to modify microscopic spin arrangements. 
Returning to the developing magnetic model of Sr$_3$NiIrO$_6$, we qualitatively expect that the ultra-high coercive field results from the high barrier to flipping a chain of strongly-coupled Ni$^{2+}$ and Ir$^{4+}$ magnetic moments, which is effectively a many-body magnetic system. This barrier to flipping results from the interaction of the magnetic order with the lattice. This interaction has two parts: the interaction of the magnetic exchange interactions with lattice distortions, and the mixing %interaction 
of the local spin-orbit coupled state of the Ir$^{4+}$ ion with the lattice. We quantified both in this work, thereby providing necessary insight to develop a complete model.

%%%%%%%%%%%%%%%%%%%%%%%%%%%%%%%%
\section*{Methods}
%%%%%%%%%%%%%%%%%%%%%%%%%%%%%%%%

High quality single crystals were grown as described previously~\cite{Singleton2016,Nguyen1995} and either polished or combined with a transparent matrix to control optical density due to strong phonon absorption. Absorption was calculated as $\alpha=\frac{-1}{hd}\ln(\cal{T}(\omega))$, where $h$ is concentration, $d$ is thickness, and $\cal{T}(\omega)$ is the measured transmittance. Magneto-infrared and magnetization measurements were carried out at the National High Magnetic Field Laboratory using the 35~T resistive and 65~T short-pulse magnets, respectively. Absorption was obtained at zero field, and magneto-infrared measurements tracked changes. Absorption differences are calculated as $\Delta\alpha=\alpha($B$)-\alpha(0$~T). We integrate $|\Delta\alpha|$ over small energy windows (127-150, 290-335, and 530-545~cm$^{-1}$) to quantify changes, and renormalize to the 35~T magnetization squared to match energy scales and provide a proper comparison.

First principles calculations were performed using projector augmented waves as implemented in VASP~\cite{Kresse1996,Kresse1996a}. The PBEsol functional was used to approximate the exchange correlation energy~\cite{Perdew1996,Perdew2008}, and DFT+$U$ %implementation due to Liechtenstein {\it et al}.~ 
\cite{Liechtenstein1995} was utilized for the transition metals. For calculations of Sr$_3$NiIrO6, $U$=5~eV for Ni and $U$=1~eV for Ir were used. For the Cu analog, $U$=5~eV for Cu and $U$=1.5 for Ir were used. Changing the values of $U$ and $J$ changes the obtained phonon frequencies but does not affect the physical picture. Phonons were calculated using both the direct method and density functional perturbation theory, and the results were identical within numerical noise. An $8\times8\times8$ Monkhorst-Pack grid \cite{Monkhorst1976} and a plane wave energy cutoff of 500~eV gave good convergence. The crystal structure used in the phonon calculations was obtained by relaxing the experimentally reported structure in the magnetically ordered phase.

The complexity of the crystal and magnetic structure of Sr$_3$NiIrO$_6$ makes it impractical to perform a detailed first-principles calculation of the spin-phonon or spin-lattice coupling in this material. In order to gain insight about the spin-lattice coupling strength, we therefore performed noncollinear magnetic calculations in a series of crystal structures that are obtained by displacing the atoms according to the displacement pattern of particular phonon modes. (Results are presented in Fig. 1 (j).) We considered displacements up to 0.5 \AA~in total for the 22 atom unit cell, which corresponds to an average of $\sim 0.1$ \AA~per atom. The magnetic ground state does not change, and is still predominantly ferrimagnetic along the $c$ axis for all of these structures. However, unlike the collinear ferrimagnetic state observed in the experiment and reproduced by first principles calculations, the magnetic ground state for these displaced structures are tilted ferrimagnetic. The 310 cm$^{-1}$ displacement gives rise to large tilting of Ni moments - as much as $\approx$10$^{\circ}$ for the larger displacements we considered. For reference, we present also the results of the same calculation for the 177~cm$^{-1}$ mode of same symmetry, and find a much smaller tilting. This observation is not surprising, because the character of this mode does not change the exchange pathways significantly. (It is of 49\% Sr, 13\% Ni, 8\% Ir, and only 29\% O character.) The finding is also in line with the experimental result that this mode is not engaged in spin-lattice coupling.

%%%%%%%%%%%%%%%%%%%%%%%%%%%%%%%%
\section*{Data Availability}
%%%%%%%%%%%%%%%%%%%%%%%%%%%%%%%%

Data is available from the corresponding author upon reasonable request.

%%%%%%%%%%%%%%%%%%%%%%%%%%%%%%%%
\section*{Acknowledgments}
%%%%%%%%%%%%%%%%%%%%%%%%%%%%%%%%

Research at the University of Tennessee, Rutgers University, and University of Minnesota is supported by the National Science Foundation DMREF program (DMR-1629079, DMR-1629059, and DMR-1629260). The crystal growth was partially supported by the National Research Foundation of Korea (NRF) funded by the Ministry of Science and ICT (No. 2016K1A4A4A01922028). We also appreciate  funding from the U.S. Department of Energy, Basic Energy Sciences, contract DE-FG02-01ER45885 (Tennessee), ``Science at 100 Tesla" (LANL), and ``Topological phases of quantum matter and decoherence" (LANL). The NHMFL facility is supported by the U.S. National Science Foundation through Cooperative Grant DMR-1644779, the State of Florida, and the U.S. Department of Energy.

%%%%%%%%%%%%%%%%%%%%%%%%%%%%%%%%
\section*{Competing Interests}
%%%%%%%%%%%%%%%%%%%%%%%%%%%%%%%%

The authors declare no competing interests.

%%%%%%%%%%%%%%%%%%%%%%%%%%%%%%%%
\section*{Author Contributions}
%%%%%%%%%%%%%%%%%%%%%%%%%%%%%%%%

J.L.M., T.B. and S.-W.C. devised the project. X.L. and S.-W.C. grew the samples. K.R.O., A.A., K.D.H., A.L.B.,  M.O., and J.L.M. carried out the spectroscopic measurements. V.Z., C.V.T., and J.S. measured the magnetic properties. A.P. and T.B. performed the theoretical calculations. K.R.O., J.L.M., V.Z., and T.B. wrote the manuscript, and all authors contributed to it.

%%%%%%%%%%%%%%%%%%%%%%%%%%%%%%%%
%\begin{references}
%%%%%%%%%%%%%%%%%%%%%%%%%%%%%%%%

%\bibliographystyle{NatComm}
%\bibliography{LibraryAbbr}

%%%%%%%%%%%%%%%%%%%%%%%%%%%%%%%%
%\section*{Additional Information}
%%%%%%%%%%%%%%%%%%%%%%%%%%%%%%%%

%\subsection*{Supplementary Information}
%The accompanying Supplementary Information provides further details of vibrational mode assignments, displacement patterns, and vibronic coupling in Sr$_3$CuIrO$_6$, as well as a discussion of the full spin-phonon Hamiltonian.

%%%%%%%%%%%%%%%%%%%%%%%%%%%%%%%%
%\section{Figure Legends}
%%%%%%%%%%%%%%%%%%%%%%%%%%%%%%%%

\end{document}